\documentclass[lettersize,journal]{IEEEtran}
\usepackage{amsmath,amsfonts}
\usepackage{algorithmic}
\usepackage{algorithm}
\usepackage{array}
\usepackage[caption=false,font=normalsize,labelfont=sf,textfont=sf]{subfig}
\usepackage{textcomp}
\usepackage{stfloats}
\usepackage{url}
\usepackage{verbatim}
\usepackage{graphicx}
\usepackage{cite}
\usepackage{xcolor}
\begin{document}

\title{Can TDD Be Employed in LEO SatCom Systems?\\Challenges and Potential Approaches}

\author{Hyunwoo Lee, Ian P.~Roberts, Jehyun Heo, Joohyun Son, Hanwoong Kim, Yunseo Lee, Daesik Hong%
\thanks{H.~Lee, J.~Son, H.~Kim, Y.~Lee and D.~Hong are with the Information Telecommunication Lab (ITL), School of Electrical and Electronic Engineering, Yonsei University, Seoul, South Korea.

I.~P.~Roberts is with the Wireless Lab, Department of Electrical and Computer Engineering, UCLA, Los Angeles, CA, USA.
H.~Lee, J.~Son, and Y.~Lee are also with the Wireless Lab at UCLA.

J.~Heo is with Samsung Electronics, Suwon 16677, South Korea.
}%
}

\maketitle

\begin{abstract}
Frequency-division duplexing (FDD) remains the de facto standard in modern low Earth orbit (LEO) satellite communication (SatCom) systems, such as SpaceX's Starlink, OneWeb, and Amazon's Project Kuiper.
While time-division duplexing (TDD) is often regarded as superior in today's terrestrial networks, its viability in future LEO SatCom systems remains unclear. 
This article details how the long propagation delays and high orbital velocities exhibited by LEO SatCom systems impedes the adoption of TDD, due to challenges involving the frame structure and synchronization.
We then present potential approaches to overcome these challenges, which vary in terms of resource efficiency and operational/device complexity and thus would likely be application-specific.
We conclude by assessing the performance of these proposed approaches, putting into perspective the tradeoff between complexity and performance gains over FDD.
Overall, this article aims to motivate future investigation into the prospects of TDD in LEO SatCom systems and solutions to enable such, with the goal of enhancing future systems and unifying them with terrestrial networks.
\end{abstract}


\section{Introduction}\label{sec_1}
By the day, low Earth orbit (LEO) satellite communication (SatCom) systems continue to cement themselves as key components of the 6G landscape, with sights set on providing near-global wireless coverage~\cite{LEO_magazine}.
Today, LEO SatCom systems such as SpaceX's Starlink, OneWeb, and Amazon's emerging Project Kuiper have become a newfound source of broadband connectivity for millions of un-/under-served users across the globe. 
These LEO SatCom systems have made tremendous strides toward the dream of ubiquitous connectivity in providing high-speed Internet access to remote areas and supporting emergency SOS and text messaging directly to/from mobile handsets. 
The continued rollout of LEO SatCom systems continues to accelerate and suggests that their use cases and requirements will diversify greatly over the coming years, with expanded support of Internet-of-things (IoT) services, connected/autonomous vehicles, and national security missions across the globe.
It remains an open question how to best optimize and deploy future LEO SatCom systems for these emerging use cases and how such systems may be integrated into the existing terrestrial network paradigm.

In any given wireless communication system, the decision to separate downlink (DL) and uplink (UL) either in time or in frequency may at first glance seem arbitrary, as both divide the time-frequency resource plane in one dimension or the other.
As the requirements and techniques of wireless communication systems continue to evolve, however, the question of whether to employ frequency-division duplexing (FDD) or time-division duplexing (TDD) has become more nuanced than ever before.
Below, we briefly summarize the principal ways in which FDD and TDD are often compared: 
\begin{itemize}
    \item \textbf{Guard resources:} 
    In both FDD and TDD, empty guard resources are required to sufficiently isolate DL and UL signals. 
    In FDD, an empty guard band is often necessary between the DL frequency band and UL frequency band to reduce adjacent channel interference.
    In TDD, an empty guard period is necessary between a DL time slot and an UL time slot to avoid collisions caused by propagation delays and accommodate the timing advances needed to compensate for said delays.
    This will be particularly relevant in the context of LEO SatCom, as we will see.
    \item \textbf{Resource allocation:}
    In most FDD systems, the DL and UL frequency bands are fixed in both center frequency and bandwidth, often based on a regulated spectrum allocation.
    As a consequence, the ratio of resources a system allocates to DL and UL is also fixed, constraining its ability to adapt to changes in DL/UL traffic demand.
    TDD systems, on the other hand, are free to allocate resources dynamically according to DL/UL demand by varying the fraction of time slots given to each. 
    \item \textbf{CSI acquisition:}
    Obtaining channel state information (CSI) is essential to reliable, high-rate communication, with both FDD and TDD.
    In FDD, since DL and UL operate within different (potentially quite distant) frequency bands, it is necessary to feed back CSI estimates from the receiver to the respective transmitter in each band.
    TDD famously circumvents the overhead, delay, and quantization associated with this feedback process by exploiting channel reciprocity, i.e., the fact that DL and UL traverse across (approximately) the same channel.
    \item \textbf{Hardware:}
    In FDD, since transmission and reception take place simultaneously at a given device, a single antenna with a duplexer or separate antennas for DL and UL is required.
    In TDD, a single antenna with a switch can be used, since DL and UL share the same frequency band yet occur in different time slots at any given device.
\end{itemize}

\newpage

For several good reasons, FDD has been the de facto standard in LEO SatCom systems deployed today and adopted by regulators across the globe in how they allocate spectrum to such systems.
We begin the remainder of this article by articulating reasons why TDD has yet to be employed in SatCom systems, highlighting how the defining characteristics of LEO---namely the long propagation delay and high Doppler---complicate its adoption.
We then discuss the potential merits of TDD in LEO SatCom, despite its associated challenges, and present plausible approaches to overcome these challenges and enable future TDD-based LEO SatCom systems.
Finally, we conclude by assessing the performance and tradeoffs of each proposed approach in comparison to FDD.

\section{Why Is TDD Not Used In LEO SatCom Today?}\label{sec_3}
In terrestrial networks, both FDD and TDD have been employed widely, with TDD adopted more often in recent years, especially in mid-bands, to take capitalize on its aforementioned advantages.
In LEO SatCom systems, however, FDD has been the predominant choice, despite the various advantages of TDD. 
To fully understand why this is the case, it is essential that we first highlight the defining characteristics of LEO SatCom systems, contrasting them with their terrestrial counterparts, and then detail the challenges these characteristics introduce in the context of TDD.

\subsection{Defining Characteristics of LEO SatCom}
Perhaps the two most noteworthy characteristics of LEO SatCom, when compared to terrestrial communication systems, are (i) the long distances over which SatCom systems communicate and (ii) the high mobility of LEO satellites.
As we will see, these two defining characteristics play a central role in impeding the adoption of TDD in LEO SatCom.

\textbf{Long communication distances:}
Typical LEO SatCom systems today orbit the Earth at altitudes ranging from 300\,km to 1,500\,km, and a single satellite may serve ground user equipments (UEs) across a vast coverage footprint whose radius is on the order of 1,000\,km.
As a representative example, let us consider a LEO satellite at an altitude of 600\,km, in line with those deployed in SpaceX's Starlink constellation, for instance, and a minimum service elevation angle of 10 degrees.
Such a satellite may communicate with UEs on the surface of the Earth over distances ranging from 600\,km to around 2,000\,km~\cite{38.811}. 
These long distances introduce two notable physical effects.
The first is a long propagation delay of at least 2\,ms, which is thousands of times greater than the maximum propagation delay seen in terrestrial cellular networks of 0.9\,us, based on a worst-case UE at the edge of an urban macrocell~\cite{LEO_MIMO_survey}.
The second effect is a potentially large differential delay between any two UEs across the coverage area. 
This is because the difference in the distance between the satellite and any two UEs can be significantly large due to the extensive coverage footprint of LEO SatCom systems. 
This leads to a differential delay of up to 4.44\,ms between UEs, based on our running example. 

\textbf{High orbital velocities:}
LEO satellites orbit the globe at velocities around 7.12-7.73\,km/s, depending on their altitude~\cite{38.811}, orders of magnitude faster than velocities experienced in terrestrial networks.
This extreme mobility induces two effects.
The first effect is a substantial Doppler shift due to satellite velocity, which is compounded by the fact that LEO SatCom systems mostly operate at relatively high carrier frequencies of 10--30\,GHz.
At these high frequencies, the Doppler frequency shift can reach up to 681\,kHz~\cite{38.811}. 
In addition to a Doppler shift, some Doppler spreading also occurs due to the time-varying nature of the channel, but this is expected to be fairly minimal due to the presence of few scatterers in the channel~\cite{LEO_MIMO}.
The second effect is time-variability in propagation delay and in Doppler shift.
Because a LEO satellite is constantly moving relative to the UEs it serves, the propagation delay between the satellite and those UEs, along with the Doppler shift induced, varies throughout the satellite's orbit.
The high orbital velocity, along with the long communication distance described before, introduce two noteworthy challenges that complicate the adoption of TDD in LEO SatCom systems, as detailed next.

\begin{figure}[!b]
\centering
\includegraphics[width=1\columnwidth]{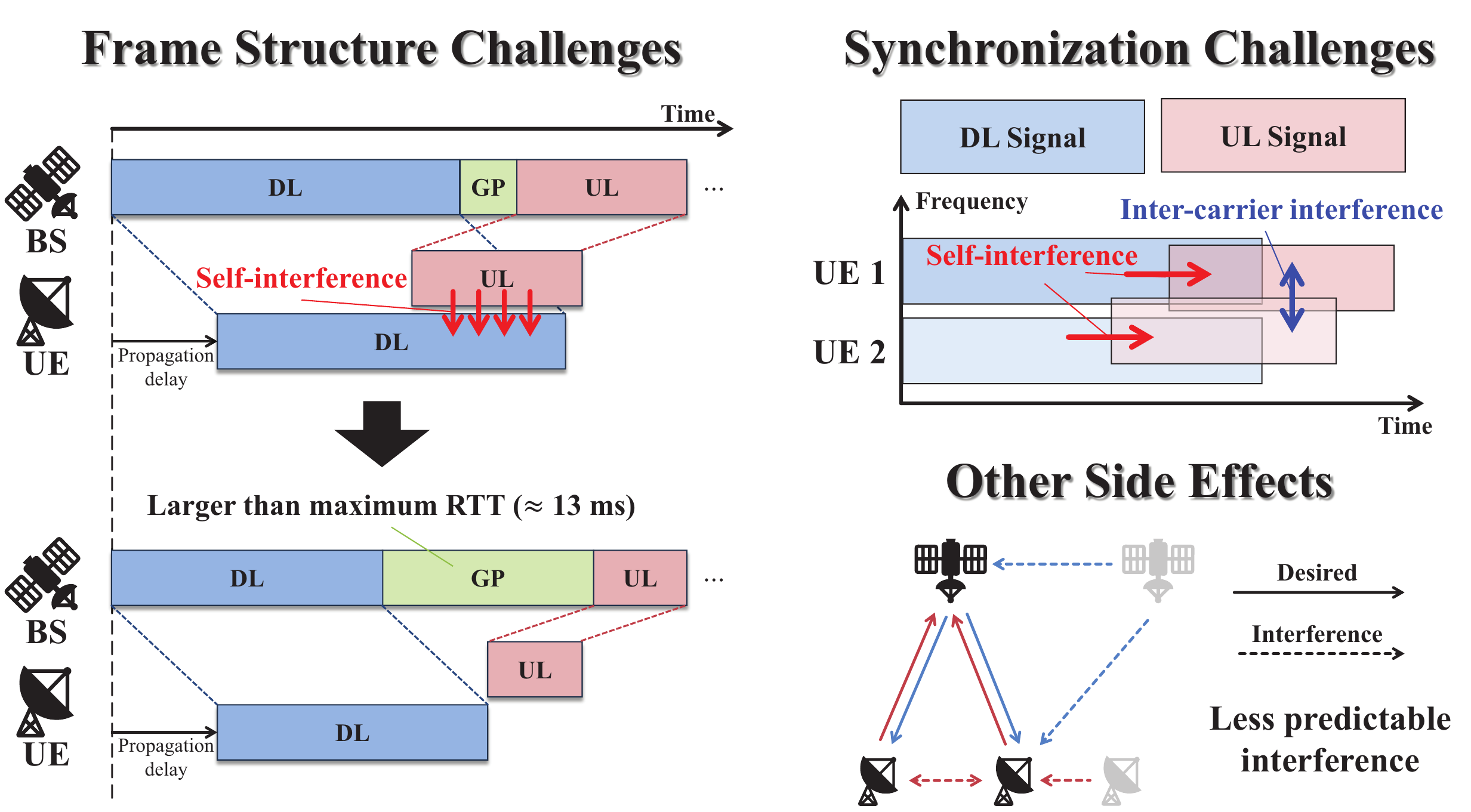}
\caption{Challenges associated with employing TDD in LEO SatCom systems. GP indicates guard period.}
\label{fig_LEO_TDD_Challenges}
\end{figure}

\subsection{Challenge \#1: Frame Structure}

The first challenge faced by the adoption of TDD in LEO SatCom systems pertains to complications in the signaling frame structure, due to the long propagation delays when communicating over distances on the order of hundreds of kilometers.
Every TDD system, including terrestrial ones, rely on an empty guard period to prevent DL and UL signals from overlapping in time, due to propagation delays in the wireless channel---a wasteful yet necessary use of radio resources. 
In terrestrial cellular networks, the TDD frame configuration is base station centric, in the sense that the timing of frames is based on when they depart or arrive at the base station, not the UEs~\cite{38.211}.
Consequently, on the UL, UEs must transmit their signals in advance to ensure they arrive at the base station on time, and a guard period is thus needed between DL and UL time slots to accommodate these timing advances.
Since the timing advance applied by each UE depends on the propagation delay of its particular channel, the guard period must be long enough to accommodate worst-case UEs with the longest propagation delay.
More specifically, as illustrated in the left side of Fig.~\ref{fig_LEO_TDD_Challenges}, the guard period must be at least twice the maximum propagation delay to allow all UEs to receive DL signals and then apply their timing advances upon transmitting UL~\cite{Flexible}.
In terrestrial networks employing TDD, this results in a guard period of around 1.8\,us~\cite{LEO_MIMO_survey}.
If TDD were used in LEO SatCom systems, the guard period must be at least around 13\,ms to accommodate a maximum propagation delay of 6.44\,ms, for our running example.
This 13\,ms guard period is thousands of times greater than that of terrestrial cellular networks and, all else being equal, amounts to a thousands-fold increase in wasted radio resources.
In Section~\ref{sec_5}, we introduce three plausible approaches to overcome this waste in resources through a redesign of the frame structure.

\subsection{Challenge \#2: Timing and Frequency Synchronization}

Synchronization in both time and frequency is essential to virtually any modern communication system.
In the context of TDD, timing synchronization is responsible for aligning DL and UL signals in their designated time slots, while frequency synchronization is necessary to ensure that transmitted and received signals are at the designated frequency.
If timing and frequency synchronization are not performed properly, inter-user and inter-carrier interference of various forms can manifest.
For instance, if UEs do not synchronize correctly in time, their UL signals will reach the base station misaligned, losing orthogonality in time and thus in frequency when orthogonal frequency division multiplexing (OFDM) is employed.
The main challenge faced by LEO SatCom systems in terms of precise timing and frequency synchronization lies in the fact that timing and frequency offsets are more extreme and more dynamic than in terrestrial networks.

In regards to timing synchronization, the large differences in propagation delay across UEs served by a single LEO satellite leads to an excessively long waiting time for timing synchronization, since this involves receiving signals from \textit{all} UEs attempting to connect~\cite{38.211}.
Recall, the propagation delay difference between the closest UE and the farthest UE from the satellite may be up to 4.44\,ms, meaning this amount of time is required during synchronization periods to ensure signals from all UEs are received.
This waiting time is nearly as long as the entire frame length in terrestrial networks~\cite{38.211}, resulting in significant resource inefficiency, if not remedied. 

In regards to frequency synchronization, the most notable challenge is the need to correct high Doppler frequency offsets, stemming from the high velocity of satellites and relatively high carrier frequencies of around 10--30\,GHz.
As mentioned, frequency offsets in LEO SatCom can be over 600\,kHz, which is more than 20 times larger than the subcarrier spacing of 30\,kHz used in OFDM by terrestrial networks~\cite{38.811}.
These challenges associated with timing and frequency synchronization in LEO SatCom are exacerbated by the fact that propagation delays and Doppler shifts are constantly changing as a given satellite traverses along its orbit.
As a result, the resources dedicated to timing and frequency synchronization can be prohibitively high in LEO SatCom, degrading system efficiency.
In Section~\ref{sec_6}, we introduce a potential approach to overcome these synchronization challenges more efficiently than conventional techniques used in terrestrial systems. 

\subsection{Other Side Effects of TDD in LEO SatCom Systems}
Beyond the major two challenges presented before---regarding the frame structure and synchronization---notable other side effects manifest when employing TDD in LEO SatCom systems.
Perhaps most relevant is the fact that interference is, in general, less predictable in TDD SatCom systems than in FDD SatCom systems and TDD terrestrial systems. 
In FDD SatCom systems, DL and UL interference is ever-present to some degree yet confined to their particular frequency bands across an entire network.
In TDD systems, cross-link interference between DL and UL signals across a network can lead to degraded system performance.
Terrestrial networks minimize this cross-link interference by synchronizing the timing of DL and UL across base stations, except in settings where neighboring base stations employ so-called dynamic TDD with different resource configurations~\cite{DTDD}.

In the case of TDD LEO SatCom systems, however, the propagation delay between a satellite and the UEs it serves can vary widely across satellites in a constellation.
As a result, even if synchronization is perfectly achieved between a specific satellite and its UEs, it is not necessarily straightforward to synchronize DL and UL transmissions throughout the entire network. 
Consequently, a UE receiving DL from its serving satellite may incur interference from UL signals transmitted by a nearby UE connected to a neighboring satellite.
Similarly, a satellite receiving UL signals from its UEs may incur interference from DL signals transmitted by a neighboring satellite or UL signals transmitted by other UEs. 
Furthermore, cross-link interference can even occur across UEs served by a single satellite if guard periods are not sufficiently long, since the propagation delays---and thus the aforementioned timing advances---can vary widely among UEs in the satellite's coverage area. 
While interference is seemingly inevitable in a dense LEO SatCom system, the fact that it is far less predictable and controllable when employing TDD is a noteworthy side effect that may further impede its adoption.

\begin{figure}[!b]
\centering
\includegraphics[width=1\columnwidth]{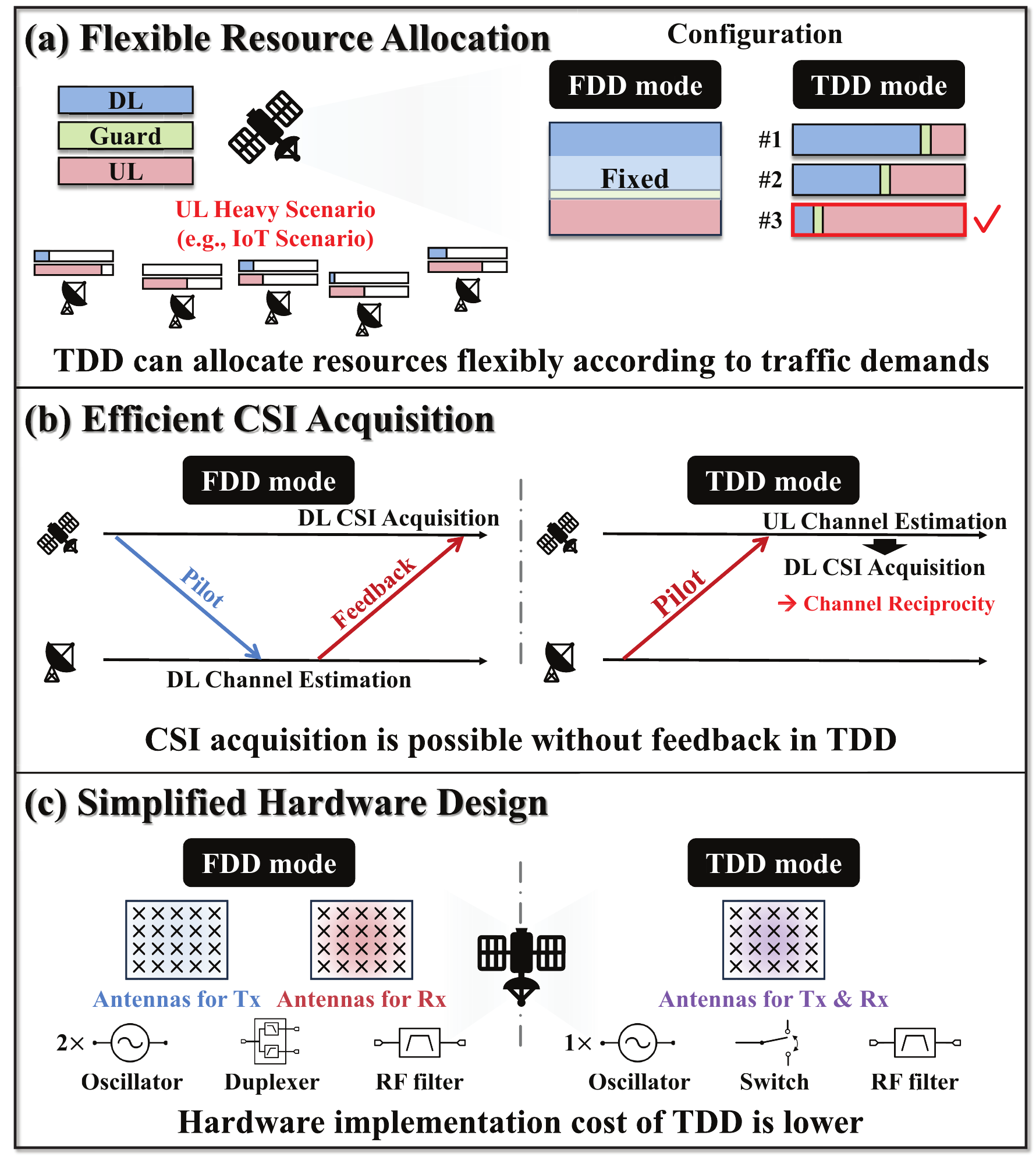}
\caption{Advantages of LEO SatCom with TDD: (a) Flexible resource allocation, (b) Efficient CSI acquisition, (c) Simplified hardware design.}
\label{fig_TDD_advantages}
\end{figure}

\section{Advantages of TDD in LEO SatCom}\label{sec_4}

The challenges associated with employing TDD laid forth thus far certainly convey why FDD has been the predominant duplexing mode adopted in LEO SatCom systems to date.
Nonetheless, in light of the widely known benefits of TDD in terrestrial networks, we now highlight three advantages TDD may offer in the context of emerging LEO SatCom networks, which we also illustrate in Fig.~\ref{fig_TDD_advantages}.

\subsection{Flexible Resource Allocation}
The role of LEO SatCom systems in future connectivity is largely dominated by two applications: (i) providing the un/under-served with broadband connectivity and (ii) serving as a backbone for remote IoT sensing and asset tracking.
LEO SatCom systems are thus faced with supporting both high DL demand and high UL demand, and this demand may vary widely over time and across the globe~\cite{BH}.
Flexible resource allocation is therefore an important component to the successful deployment of LEO SatCom systems, and the advantages of TDD over FDD in this regard may prove beneficial. 
In FDD, the frequency bands allocated to DL and UL are fixed, which restricts the ability of a system to adapt to variations in DL/UL traffic demand.
On the other hand, in TDD, the resources allocated to DL and UL can be adjusted by modifying the slot configuration~\cite{38.211}, offering better flexibility for addressing asymmetric traffic demands.
For example, as shown in Fig.~\ref{fig_TDD_advantages}(a), a LEO SatCom system employing TDD can dedicate more time slots to UL than DL to meet high UL demand, whereas a FDD system is unable to modify the DL and UL frequency bands it has been allocated.

\subsection{Efficient CSI Acquisition}
Fast yet reliable CSI acquisition is a crucial component of many modern communication systems to optimize both transmission and reception of signals.
In the context of LEO SatCom systems, long propagation delays complicate CSI acquisition when FDD is employed, as evidenced by its three-step pilot-based channel estimation process:
(1)~A pilot signal is transmitted on the DL frequency band from the satellite to the UEs.
(2)~Channel estimation is performed at each UE.
(3)~Each UE sends CSI feedback to the satellite using the UL frequency band.
The sheer distance between the satellite and its UEs leads to substantial latency in CSI acquisition and hence so-called CSI aging as the channel changes thereafter. 
Furthermore, as with terrestrial systems employing FDD, CSI quality is degraded by quantization when fed back and consumes additional radio resources that could otherwise be used for communication.
TDD, on the other hand, famously offers a simpler and more accurate CSI acquisition process, owing to channel reciprocity~\cite{FDD_prediction}.
By performing channel estimation at the satellite using UL pilot signals transmitted by the UE, the satellite can infer DL CSI directly from UL CSI through channel reciprocity.
This circumvents the need for CSI feedback, meaning CSI acquisition does not suffer from quantization and requires only half the time as FDD, which can be substantial in combating CSI aging, given the long propagation delays in LEO SatCom systems.

\begin{figure*}[!b]
\centering
\includegraphics[width=2\columnwidth]{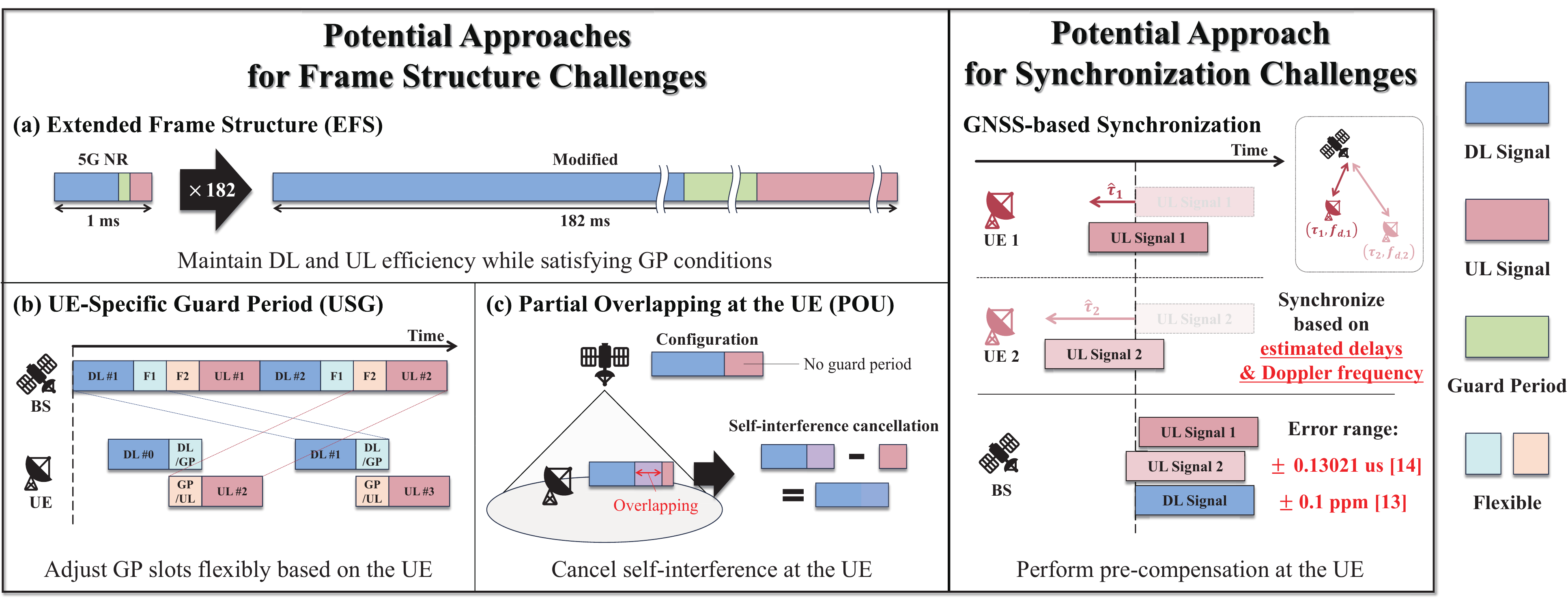}
\caption{Potential approaches for frame structure challenges: (a) Extended frame structure, (b) UE-specific guard period, (c) Partial overlapping at UE, and potential approach for synchronization challenges: GNSS-based synchronization. GP indicates guard period.}
\label{fig_potential_approach}
\end{figure*}

\subsection{Simplified Hardware Design}

TDD is often more advantageous than FDD in the sense that it can offer simplified hardware implementations.
More specifically, FDD requires simultaneous transmission and reception of DL and UL signals, and this often necessitates the use of a duplexer or physically separated antennas to isolate DL and UL.
Note that, even when DL and UL signals are transmitted and received in different frequency bands, they can interfere with each other due coupling and resonances within transceiver circuitry~\cite{RF_book}.
To prevent this interference, the transmit and receive chains of a given device must be isolated. 
This isolation can often be reliably achieved by using sharp radio frequency filters and two separate antennas or by using duplexers~\cite{RF_book}, though the latter does not scale well to the phased arrays used by satellites and UEs in modern LEO SatCom systems.
Moreover, separate local oscillators are required by FDD for transmitting and receiving DL and UL signals, as they are at different, often distant carrier frequencies.
On the other hand, a TDD transceiver requires only a single antenna and a single oscillator, as the same frequency band is used for both DL and UL. 
Although a switch is required for transitioning between DL and UL operation, this can be relatively inexpensive and smaller than other radio frequency components~\cite{switch}.
Consequently, the hardware design of TDD-based transceivers can offer lower costs, smaller form factors, and less weight than their FDD counterparts.
This can be particularly useful in deploying low-cost user terminals or lighter-payload LEO satellites.

\section{Overcoming Frame Structure Challenges}\label{sec_5}
To fully leverage the potential benefits of TDD in LEO SatCom systems, addressing its associated challenges is imperative.
As shown in Fig.~\ref{fig_potential_approach}, this section and the next suggest potential approaches to address the two key challenges presented before in Section~\ref{sec_3}.
We begin by first introducing three potential approaches to overcome the frame structure challenges in this section, leaving those for the synchronization challenges to the next section.

\subsection{Approach \#1: Extended Frame Structure}
Recall, the key challenge associated with the frame structure in TDD LEO SatCom systems is the fact that a prohibitively long guard period is required to accommodate the propagation delay and corresponding timing advance at any given UE.
This empty guard period is a necessary yet wasteful use of resources.
The first approach we introduce to rectify this is perhaps the most natural: extending the frame proportional to the guard period, as illustrated in Fig.~\ref{fig_potential_approach}(a).
In 5G NR, one time slot out of every 14 serves as a dedicated guard period~\cite{38.211}.
To attain the same level of efficiency, TDD LEO SatCom systems could extend the duration of frames by a factor of about 182, proportional to the elongated guard period.
This would ensure that time allotted to DL and UL is 14 times that of the guard period, as done in terrestrial 5G NR systems. 
While this extended frame structure serves as a straightforward way to improve the resource efficiency in TDD LEO SatCom, it exhibits some noteworthy practical drawbacks, including increased latency and severe CSI aging, even when exploiting channel reciprocity, due to the long frame length. 
We address these drawbacks in the next two approaches.

\subsection{Approach \#2: UE-Specific Guard Periods}
Rather than merely extend the frame structure, a second plausible approach is to make the guard period duration UE-specific.
As described before, resource inefficiency arises when the guard period duration is based on the UE with the longest propagation delay to ensure all serviced UEs are free from overlapping of DL and UL.
Instead, it is in fact viable to employ \textit{UE-specific} guard periods, as illustrated in Fig.~\ref{fig_potential_approach}(b).
Here, if no guard period is allocated, the overlap between DL and UL signals will differ for each UE. 
Since the end of each DL signal (e.g., DL slot \#0 in Fig.~\ref{fig_potential_approach}(b)) overlaps with the start of each UL signal (e.g., UL slot \#2), the UE would incur self-interference in its attempt to receive UL. 
As illustrated, however, a guard period can be appended to the end of the DL slot or inserted at the beginning of the UL slot.
This prevents self-interference (overlapping) at the UE and does so using a guard period only as long as strictly necessary, since it is specific to the UE being served---thereby improving resource efficiency.
This concept was first introduced in a 3GPP Rel.~16 document~\cite{Flexible} and defines the overlapping portion as a flexible slot dynamically allocated to DL, UL, or guard period based on traffic demand.
Note that this approach is not practically viable in terrestrial networks, since the difference in the minimum required guard period for each UE is negligible, but is effective for LEO SatCom Systems, where it is substantially larger.

\subsection{Approach \#3: Partial Overlapping of DL/UL at the UEs}
The previous two approaches involve the use of guard periods to completely eliminate the potential overlapping of DL and UL signals at the satellite and at all UEs.
In contrast, our third approach proposes partial overlapping of DL and UL signals on the UE side, with no guard period applied by the satellite.
Suppose no guard period is used at all, in which case transmission of UL signals by a given UE must partially overlap its reception of DL signals from the satellite to accommodate the aforementioned timing advances.
Normally, this overlapping of DL and UL would substantially impede the UE's reception of its DL signal, as the UL transmission would inflict so-called self-interference.
However, if the UE was equipped with in-band full-duplex capability~\cite{SIC_peformance}, this would not be a problem, as it could transmit and receive at the same time and same frequency without penalty.
Successfully realizing such requires sufficient cancellation to rid the desired DL signal of self-interference. 
Fortunately, state-of-the-art self-interference cancellation (SIC) techniques, often employing the combination of analog and digital filtering, are capable of impressive levels of cancellation up to 130\,dB~\cite{SIC_peformance}, which would likely be sufficient for most UEs.
The effectiveness of such techniques is even more optimistic when one considers the fact that the self-interference channel would presumably be fairly static, given UEs are steered toward the sky, in view of few time-varying scatterers.
Adopting such SIC techniques to enable full-duplex UEs for this purpose would perhaps be the most straightforward way of overcoming the frame structure challenges in TDD LEO SatCom systems, as it would require little to no overhauling of the DL and UL framing---but this comes at the cost of higher UE complexity.

\section{Overcoming Synchronization Challenges}\label{sec_6}
We now turn our attention to addressing the challenge of timing and frequency synchronization, as highlighted before, which recall is rooted in the overhead that would be consumed to continuously track the timing and frequency offsets seen by each UE relative their serving satellite.
It not unreasonable to assume that each satellite in a constellation continuously broadcasts ephemeris information to UEs in modern LEO SatCom systems~\cite{38.821}. 
With this information, each UE can infer with reasonable accuracy the location and velocity of its serving satellite at virtually any time, as well as the satellite's orbital information.
It is also not unreasonable to assume that each UE is equipped with global navigation satellite system (GNSS) functionality, allowing it to infer its own location on Earth.
With these pieces of information, a given UE can estimate its propagation delay based on its calculated distance from the satellite and can thus synchronize the timing of its UL signal by applying a timing advance commensurate with this propagation delay.
Furthermore, with knowledge of the satellite's velocity and its own location, each UE can calculate the relative velocity of the satellite and can thus estimate the corresponding Doppler shift, thereby enabling pre-/post-compensation and facilitating frequency synchronization.
Together, this GNSS-based approach could provide timing synchronization within an error margin of $\pm$0.13\,us \cite{38.133} and frequency synchronization within an error margin of $\pm$0.1\,ppm~\cite{38.821}---both of which are within the TDD requirements of terrestrial networks~\cite{36.104}.

\section{Performance Analysis}\label{sec_7}
Having highlighted the enhancements offered by TDD and potential approaches to realize such, this section aims to gauge the performance of these approaches against FDD through extensive simulation of a LEO SatCom system.
To do so, we measure the so-called DL resource efficiency ratio, defined as the ratio between the spectral efficiency attained with TDD versus that with FDD, across all UEs.
The total time-frequency resources are kept equal in both TDD and FDD, for a fair comparison.
We have focused specifically on measuring DL performance, since it is more critically affected in TDD due to the partial overlap between DL and UL signals on the UE side.
Simulation parameters such as the satellite's equivalent isotropic radiated power density, satellite transmit antenna gain, UE receive antenna gain, and UE transmit power can be found in Table~6.1.1.1 of 3GPP TR 38.821~\cite{38.821}; all other relevant details are described below.
The altitude of the satellite is set to 600\,km, the minimum elevation angle to 10 degrees, the DL carrier frequency to 20\,GHz, and the bandwidth to 400\,MHz~\cite{38.811}. 
UEs are uniformly distributed over the Earth's surface within the coverage region of the satellite, whose radius is 1,761\,km.
A time-varying channel is simulated using Jakes' model, and the acquired CSI is used throughout the frame duration.
The ratio of DL-to-UL is set to 7:3 in both FDD and TDD, and the guard band for FDD is set to 5\% of the total bandwidth. 
The frame length for the extended frame structure solution is set to 182\,ms to preserve the same guard period ratio as in 5G NR, and in all other cases, the frame length is set to 1\,ms, the same as in 5G NR.
GNSS-based synchronization is assumed to yield a maximum timing synchronization error of $\pm$0.13\,us~\cite{38.133} and frequency synchronization error of $\pm$0.1\,ppm~\cite{38.821}.

\begin{figure}[!b]
\centering
\includegraphics[width=1\columnwidth]{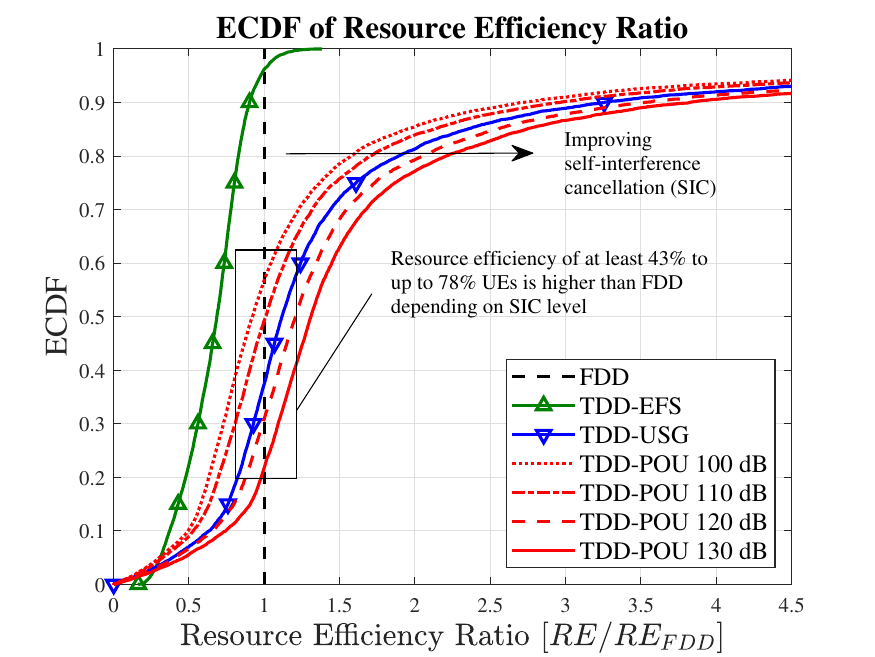}
\caption{Empirical CDF of the gain in DL resource efficiency over FDD when employing the three TDD schemes of Section~\ref{sec_5}.}
\label{fig_resource_efficiency}
\end{figure}

In Fig.~\ref{fig_resource_efficiency}, we plot the empirical CDF (across randomly dropped UEs) of the DL resource efficiency ratio for various schemes, with a ratio of one corresponding to performance with FDD.
The green line indicates performance of TDD with the extended frame structure (TDD-EFS) approach, and the blue line is that with a UE-specific guard period (TDD-USG).
The four red lines represent performance with partial overlapping at the UE for various levels of SIC (TDD-POU).
We can observe that the extended frame structure performs the poorest in terms of resource efficiency, falling short of FDD with high probability.
This can be attributed to severe CSI aging during the elongated DL and UL time slots, which were extended by a factor of 182 compared to those with FDD.
Performance is substantially improved when employing a UE-specific guard period, as around 63\% of UEs enjoy higher resource efficiency with TDD compared to FDD.
This is thanks to reduced CSI aging and fewer guard resources, compared to the extended frame structure. 
For the partial overlapping scheme, performance is comparable to that with a UE-specific guard period with modest levels of SIC.
Even with relatively low SIC of 100\,dB, the resource efficiency is superior than FDD for around 43\% of UEs.
As SIC is improved to its state-of-the-art level of 130\,dB~\cite{SIC_peformance}, this fraction grows to 78\% of UEs, a substantial improvement beyond what can be attained with the UE-specific guard period.

Table~\ref{tab_summary} serves as a concise summary of these three TDD-based techniques when compared against FDD.
First, in the case of the extended frame structure (EFS), the resource efficiency is relatively poor, but the required complexity at both the satellite and UE is low.
Recall, the poor resource efficiency is due to degraded CSI accuracy in time-varying channels (i.e., CSI aging). 
Additional pilots inserted between DL slots to continuously track the channel and/or channel prediction techniques could be used to rectify this channel aging issue at the cost of additional pilot overhead and/or computational complexity.
Next, in the case of the UE-specific guard period (USG), resource efficiency is relatively high and UE complexity is low.
This comes at the cost of increased satellite complexity, as resources need to be allocated flexibly to service UEs without overlapping DL and UL, and this requires the satellite to know the relative propagation delay to each UE.
Lastly, in the case of partial overlapping at the UE (POU), resource efficiency is relatively high, and there is little no additional complexity required at the satellite.
This comes at the cost of additional UE complexity, since they must employ SIC techniques to simultaneously transmit UL while receiving DL when such signals overlap.

Clearly, each scheme has certain quantitative and qualitative advantages and disadvantages over the others, and the appropriate scheme may thus heavily depend on use case.
The extended frame structure may be most suitable for applications such as text messaging and IoT, where the focus is not on high-speed data transmission or low latency.
The UE-specific guard period is perhaps most suitable for services targeting handheld devices, where high-speed data transmission and a low UE complexity are prioritized. 
Partial overlapping at the UE, on the other hand, is more suited for scenarios targeting larger form-factor ground terminals, where a higher UE complexity may be tolerable.

\begin{table}[!t]
\centering
\caption{Comparison of the approaches presented herein.}
\includegraphics[width=1\columnwidth]{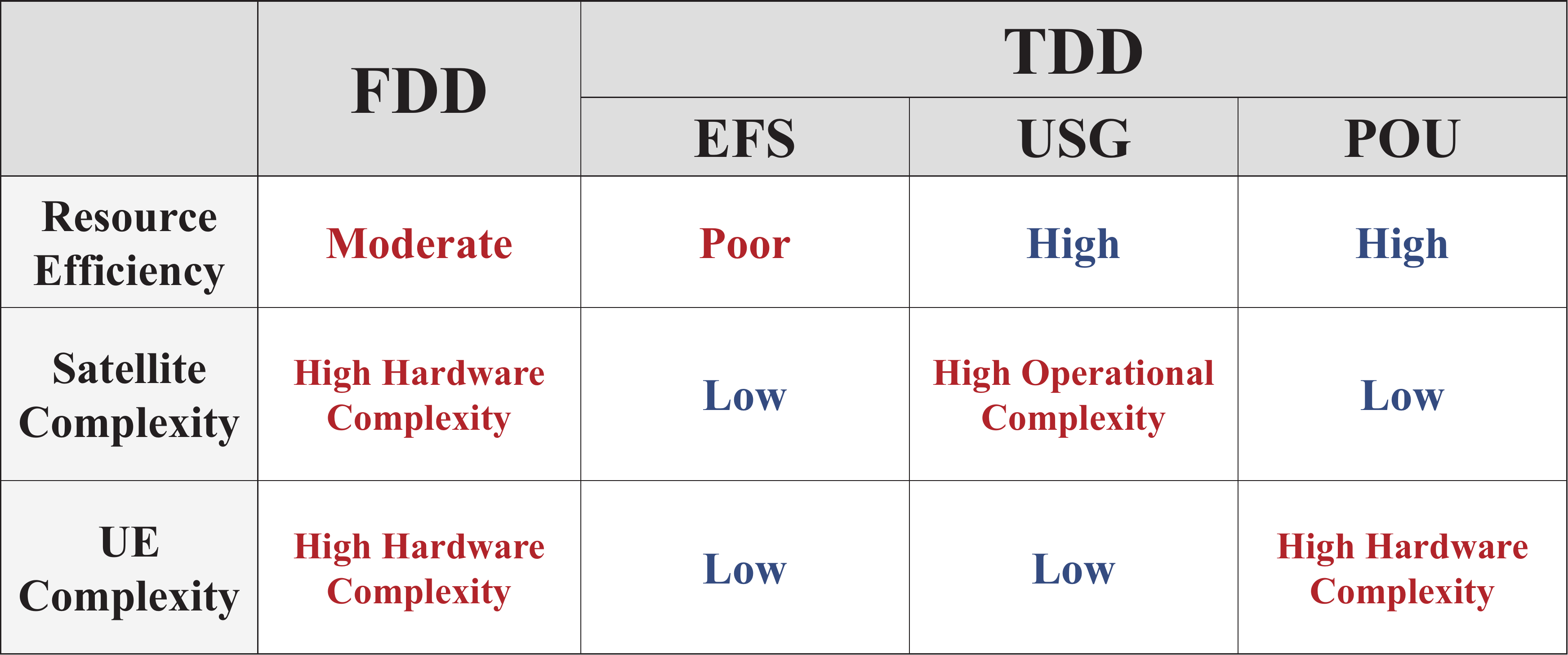}
\label{tab_summary}
\end{table}

\section{Conclusion}\label{sec_8}
Virtually all LEO SatCom systems have been deployed using FDD to separate DL and UL in the frequency domain, and this is largely due to the challenges that would arise if TDD was adopted in such systems.
Perhaps the two most notable of these challenges involve the frame structure and synchronization, which stem from long propagation delays, high Doppler shifts, and the fact that both vary widely across UEs and over time as a satellite orbits the globe.
In this article, we have introduced a variety of potential approaches to overcome these frame structure challenges through the use of an extended frame structure, a UE-specific guard period, or partial overlapping at the UE, with the latter enabled by SIC techniques.
We also highlighted how GNSS functionality at the UE can be harnessed to overcome timing and frequency synchronization challenges.
Together, these approaches show promise in deploying TDD-based LEO SatCom systems, and thus pave the way toward higher-throughput, more adaptive, and fully unified LEO networks in the 6G era.

Nonetheless, several areas of research are of importance to further explore and advance the potentials of TDD-based LEO SatCom systems, including:
\begin{itemize}
\item{Channel prediction techniques harnessing machine learning to reduce CSI acquisition overhead and combat CSI aging.}
\item{Scheduling and resource allocation algorithms when employing a UE-specific guard period.}
\item{Studying self-interference and subsequently developing SIC techniques in UE terminals, especially for use in the partial overlapping scheme presented herein.}
\item{Methods to achieve accurate timing and frequency synchronization for UEs without GNSS functionality or under severe GNSS error.}
\item{Mechanisms aimed at resolving the inter-satellite interference that arises when many satellites within a constellation employ TDD in an asynchronous manner.}
\end{itemize}

\bibliographystyle{bibtex/IEEEtran}
\bibliography{bibtex/IEEEabrv,reference}

\end{document}